\documentclass[final,5p,times,twocolumn]{elsarticle}

\usepackage{amssymb}

\usepackage{color}
\usepackage{amsmath}
\usepackage{multirow}
\usepackage{booktabs}
\usepackage{bm}
\usepackage{braket}
\usepackage{amsmath}
\usepackage{graphicx}
\usepackage{soul}
\biboptions{sort&compress}

\usepackage[pdfstartview=FitH, colorlinks,
 linkcolor={red!60!black!},
 anchorcolor={green!80!black!},
 urlcolor={blue!80!black!},
 citecolor={blue!50!black!}]{hyperref}
\usepackage[table]{xcolor}

\journal{Physics Letters B}

\begin{document}


\begin{frontmatter}



\title{The roles of three-nucleon force and continuum coupling in mirror symmetry breaking of oxygen mass region}

\author{S. Zhang}
\author{Y. Z. Ma}
\author{J. G. Li}
\author{B. S. Hu}
\author{Q. Yuan}
\author{Z. H. Cheng}
\author{F. R. Xu\corref{cor1}}

\address{School of Physics, and State Key Laboratory of Nuclear Physics and Technology, Peking University, Beijing 100871, China}

\cortext[cor1]{frxu@pku.edu.cn}

\begin{abstract}
With both three-nucleon force and continuum coupling included, we have developed a self-consistent {\it ab initio} Gamow shell model within the Gamow Hartree-Fock (GHF) basis obtained by the realistic interaction itself. With the chiral two-nucleon N$^3$LO and three-nucleon N$^2$LO interactions, the Gamow shell model has been applied to the mirror systems of $Z=8$ neutron-rich isotopes and $N=8$ proton-rich isotones, giving good agreements with data in binding energies, dripline positions and excitation spectra. The GHF calculated that the $0d_{3/2}$, $1s_{1/2}$ and $1p_{3/2}$ orbitals are resonances. The resonance states and their interplay with nonresonant continua play a crucial role in the descriptions of nuclei around driplines. Excitation spectra and Thomas-Ehrman shifts observed can be better described when both three-nucleon force and continuum coupling are considered in calculations. The three-nucleon force and continuum coupling produce a combined effect on the Thomas-Ehrman shift, e.g., for the ${1/2}^+$ resonance level of $^{19}$Na. The calculations help the understandings of related nuclear astrophysical processes.
\end{abstract}

\begin{keyword}
Three-nucleon force \sep Continuum coupling \sep Gamow shell model \sep Thomas-Ehrman shift \sep Mirror nuclei \sep Oxygen mass region
\end{keyword}

\end{frontmatter}



\section{Introduction}
The approximate charge independence of the strong interaction results in a good symmetry in excitation spectra between mirror nuclei. However, it was found that in some mirror nuclei the symmetry is broken significantly. Historically, the mirror symmetry breaking was considered to arise from the Coulomb energy purely. The difference in Coulomb energies between the excited state and the ground state (g.s.) results in different shifts of energy levels, called the Thomas-Ehrman shift (TES)~\cite{PhysRev.88.1109,PhysRev.81.412}. The change in Coulomb energy is due to the change in the spatial asymptotic behavior of nuclear states~\cite{PhysRev.88.1109,PhysRev.81.412}. Theoretical investigations~\cite{OGAWA1999157,PhysRevC.89.044327} commented that the residual nucleon-nucleon (NN) interaction may also contribute due to the extension of wave functions. Indeed, the profound reordering of levels in the mirror pair $^{16}\text{N}$-$^{16}\text{F}$ was observed, and explained by taking into account the continuum effect~\cite{PhysRevC.90.014307}. 

In the historical example of the $^{13}$C-$^{13}$N mirror pair where TES was observed~\cite{PhysRev.88.1109,PhysRev.81.412}, the first excited state in $^{13}$N is an unbound resonant state above the threshold of the $^{13}$N proton emission. Therefore, when discussing the TES, the accurate treatment of asymptotic behaviors of wave functions of weakly-bound and unbound nuclei is crucial. The Gamow shell model (GSM)~\cite{PhysRevLett.89.042501,PhysRevLett.89.042502} is a powerful tool to describe the asymptotic behavior of wave functions. It presents the coupling to the continuum at basis level by using the complex-energy Berggren basis~\cite{BERGGREN1968265}, and many-body internucleon correlations occur via configuration mixing, arising from a direct diagonalization of the complex GSM Hamiltonian. Using the Berggren basis, the complex coupled cluster~\cite{HAGEN2007169,PhysRevLett.108.242501} and the complex in-medium similarity renormalization group~\cite{PhysRevC.99.061302} have also been proposed for nuclear many-body calculations with the continuum coupling included. The GSM has been further developed with realistic two-body nucleon-nucleon (NN) interactions~\cite{PhysRevC.73.064307,PhysRevC.80.051301,PhysRevC.88.044318,SUN2017227,HU2020135206}. 

The asymptotic behavior of a nuclear state is mainly determined by the strong interaction acting between nucleons. The importance of the three-nucleon force (3NF) has been realized in nuclear structure calculations~\cite{PhysRevLett.99.042501,PhysRevLett.105.032501,PhysRevLett.106.202502,PhysRevLett.109.052501,PhysRevLett.108.242501,PhysRevLett.110.022502,PhysRevLett.110.242501,PhysRevLett.113.142501,PhysRevC.96.014303,PhysRevC.98.044305,PhysRevC.100.034324,PhysRevC.101.014318,MA2020135257,PhysRevC.102.054301,MA2020135673,PhysRevLett.126.022501}. For example, 3NF can provide a repulsive contribution to binding energies in oxygen isotopes, which resolves the overbinding problem and reproduces the dripline position~\cite{PhysRevLett.105.032501,PhysRevLett.108.242501,PhysRevLett.110.242501,PhysRevLett.113.142501,MA2020135257}.
The $Z=8$ oxygen isotopic chain and its mirror-reflected $N=8$ isotonic chain provide remarkable cases of interest, at the interface of light- and medium-mass regions. The proton-magic oxygen chain is one of the best laboratories to test advanced many-body methods. $^{25,26}$O which locate beyond the neutron dripline have been found in experiment, with $^{26}$O barely unbound with two-neutron separation energy of only $-18$ keV~\cite{PhysRevLett.116.102503}. As mirror partners of oxygen isotopes, proton-rich $N=8$ isotones have drawn particular interests in nuclear structure and nuclear astrophysics. Recent experiments show novel phenomena in their masses~\cite{PhysRevLett.113.082501,XU2017312} and excitation spectra~\cite{PhysRevLett.83.45,PhysRevC.67.014308,PhysRevC.99.021301}. These nuclei play a vital role in nuclear reactions that affect stellar nucleosynthesis, such as the $^{17}\text{F}(p,\gamma)^{18}\text{Ne}$ reaction in hot CNO cycles~\cite{doi:10.1146/annurev.nucl.012809.104505} and $^{18}\text{Ne}(2p,\gamma)^{20}\text{Mg}$ reaction for bridging the waiting point in the $rp$-process~\cite{PhysRevC.51.392}. Furthermore, proton-rich $N=8$ isotones provide us with an arena to probe the mirror symmetry. The mirror asymmetry can be more significant in nuclei near driplines, because loosely-bound and unbound resonance states have large spatial spreads in the wave function.

In the GSM calculations, usually the Woods-Saxon (WS) potential was adopted to produce single-particle (s.p.) Berggren basis, with the WS parameters determined by fitting experimental s.p. energies. Then, the numerical results would depend to a certain extent on the detail of the parameterization. In the multi-shell case, such a fit is even difficult due to the lack of experimental data of cross-shell s.p. energies. To avoid the parameter dependence, the complex-energy Gamow Hartree-Fock (GHF) method~\cite{PhysRevC.73.064307,PhysRevC.88.044318,HU2020135206} was used to generate the Berggren basis. This also leads to a more self-consistent {\it ab initio} calculation because the basis is produced by the realistic interaction itself instead of a parameterized WS potential.

\section{Theoretical framework}
The intrinsic Hamiltonian of the $A$-nucleon system reads
\begin{equation}
\begin{aligned}
H=&\sum_{i=1}^{A}\left(1-\frac{1}{A}\right) \frac{\boldsymbol{p}_{i}^{2}}{2 m}+\sum_{i<j}^{A}\left(v_{ij}^{\mathrm{NN}}-\frac{\boldsymbol{p}_{i}\cdot\boldsymbol{p}_{j}}{m A}\right)+\sum_{i<j<k}^{A}v^{\mathrm{3N}}_{ijk},
\end{aligned}
\end{equation}
where $\boldsymbol{p}_{i}$ is the nucleon momentum in the laboratory, and $m$ is the mass of the nucleon. The two-nucleon (NN) interaction $v^{\mathrm{NN}}$ takes the chiral potential at next-to-next-to-next-to-leading order ($\mathrm{N^{3}LO}$) by Entem and Machleidt~\cite{PhysRevC.68.041001}, and the 3NF $v^{\mathrm{3N}}$ chooses the chiral interaction at next-to-next-to-leading order ($\mathrm{N^{2}LO}$) which was established in~\cite{PhysRevLett.99.042501}. Due to heavy demanding in computation, the inclusion of the explicit 3NF renders many-body calculations impossible. Therefore, we transform 3NF into a normal-ordered form, and neglect the residual three-body part, which indicates that a normal-ordered two-body (NO2B) approximation of the 3NF is taken. It has been shown that the NO2B 3NF can well describe the 3NF contribution in nuclear structure calculations~\cite{PhysRevLett.109.052501}. 

With the realistic interaction including the Coulomb potential, we perform the complex-momentum (complex-$k$) GHF~\cite{PhysRevC.99.061302,PhysRevC.73.064307,PhysRevC.88.044318} calculation for $^{16}$O to generate the Berggren basis which consists of bound, resonant and scattering states. $^{16}$O is chosen as the core in the present many-body GSM calculation. With the normal-ordering approximation, Hamiltonian (1) can be written as
\begin{equation}
\begin{aligned}
\hat{H}=&E_{0}+\sum_{pq}[t_{pq}+\sum\limits_{r=1}^{A}(V_{prqr}^{\mathrm{NN}}+\frac{1}{2}W_{prqr}^{2\mathrm{B}})]:\hat{a}_{p}^{\dagger} \hat{a}_{q}:\\
       &+\frac{1}{4} \sum_{pqrs} (V^{\mathrm{NN}}_{pqrs}+W^{2 \mathrm{B}}_{pqrs}) :\hat{a}_{p}^{\dagger} \hat{a}_{q}^{\dagger} \hat{a}_{s} \hat{a}_{r}:,
\end{aligned}
\end{equation}
where $\hat{a}^\dagger_p$ ( $\hat{a}_p$) is the particle creation (annihilation) operator on the $p$-th basis orbital. The operators are in normal-ordered products with respect to the chosen reference state, indicated by colons. The Hamiltonian is composed of zero-body, one-body and residual two-body parts. The zero-body term $E_0$ is a constant, describing the energy of the reference state. The one-body term includes the particle kinetic energy $t=\frac{\boldsymbol{p}^{2}}{2\mu}$ with $\mu=m/(1-\frac{1}{A})$, and the normal-ordered one-body part from NN and 3N interactions. The two-body term, $-\frac{\boldsymbol{p}_{i}\cdot\boldsymbol{p}_{j}}{m A}$, which arises from the center-of-mass (CoM) motion, is incorporated into $V^\mathrm{NN}$ in practical calculations. The normal-ordered two-body part of the 3NF is defined by
\begin{equation}
W^{2 \mathrm{B}}_{pqrs}=\sum_{h \in \mathrm{hole}}V^{3 \mathrm{N}}_{pqh,rsh},
\end{equation}
where $V^{3 \mathrm{N}}_{pqh,rsh}$ stands for antisymmetrized 3NF matrix elements. The sum runs over all hole states in the core nucleus in the HO basis. We see that the normal-ordered 3NF contributes to both one- and two-body interactions in Hamiltonian (2). 

The one-body part of Hamiltonian (2) can be considered as the motion of an independent particle in the Hartree-Fock (HF) field, with the single-particle Hamiltonian given by,
\begin{equation}
\begin{aligned}
h^{\text{HF}}_{pq}=t_{pq}+\sum\limits_{r=1}^{A}(V^{\mathrm{NN}}_{prqr}+\frac{1}{2}W^{2\mathrm{B}}_{prqr}).
\end{aligned}
\end{equation}
The HF equation can be solved in the HO basis, giving real-energy HF eigen solutions including one-body HF potential~\cite{PhysRevC.99.061302}. To describe the resonance and continuum properties of HF basis states, the HF single-particle Hamiltonian is written in the complex-$k$ space, as
\begin{equation}
\begin{aligned}
\bra{k}h\ket{k^\prime}=&\frac{\hbar^2 k^2}{2\mu}\delta(k-k^\prime)+\sum_{\alpha\beta}\bra{\alpha}U\ket{\beta}\Braket{k|\alpha} \Braket{\beta|k^\prime},
\end{aligned}
\end{equation}
where $\bra{\alpha}U\ket{\beta}$ is the one-body HF potential which is obtained by solving the real-energy HF Equation~(4), while $\Braket{k|\alpha}$ is the HO basis wavefunction $|\alpha\rangle$ expressed in the complex-$k$ space. Eq.~(5) gives the GHF basis which contains bound, resonance and continuum states. In numerical calculations, the GHF equation is solved using the Gauss-Legendre quadrature discretization scheme~\cite{PhysRevC.73.064307} in the complex-$k$ plane with a deformed contour $L^+$ which encompasses resonance states. We find that 35 discretization points are sufficient to make calculations converged~\cite{SUN2017227,HU2020135206,PhysRevC.103.034305}. Details of the GHF solution can be found in~\cite{PhysRevC.99.061302,PhysRevC.73.064307,PhysRevC.88.044318}.

Within the GHF basis obtained with choosing $^{16}$O as the reference state, we transform the chiral interaction matrix elements from the HO basis to the GHF basis for many-body GSM calculations. This can be achieved by computing overlaps between the GHF and HO basis wave functions~\cite{PhysRevC.102.034302}. In practical calculation, we take the HO basis at $h\omega =14$ MeV with 13 major shells (i.e., $e=2n+l \leq e_{\mathrm{max}}=12$) and limiting $e\leq 6$ for 3NF. To expedite the convergence of calculations, the chiral NN interaction is evolved to a low momentum scale $\lambda_{\mathrm{SRG}}=2.3$ fm$^{-1}$ using the similarity renormalization group (SRG)~\cite{PhysRevC.75.061001,BOGNER201094}. 

The GHF calculation gives that the $0d_{5/2}$ orbital is bound for both neutron ($\nu$) and proton ($\pi$), while $0d_{3/2}$ and $1p_{3/2}$ are resonances. The $\nu 1s_{1/2}$ is bound, while the $\pi 1s_{1/2}$ is a resonance. Therefore, the $d_{3/2}$, $p_{3/2}$ and $\pi s_{1/2}$ partial waves are treated in the complex-$k$ GHF basis, while all other channels are represented in the real-energy discrete HF basis to reduce the model dimension and computational task. Due to the high centrifugal barrier of the partial waves bearing $l\ge 3$, their influence on wave function asymptotes is negligible~\cite{PhysRevC.103.034305}. Therefore, the $f_{7/2}$ channel is approximated within the real-energy HF basis. Finally, the active space for the present GSM takes $\nu$$\{$$0d_{5/2}$, $1s_{1/2}$, $d_{3/2}$ resonance plus continuum, $p_{3/2}$ resonance plus continuum, $0f_{7/2}$$\}$ and $\pi$$\{$$0d_{5/2}$, $s_{1/2}$ resonance plus continuum, $d_{3/2}$ resonance plus continuum, $p_{3/2}$ resonance plus continuum, $0f_{7/2}$$\}$.

We use the many-body perturbation theory (MBPT) to construct the GSM effective Hamiltonian for the chosen model space. In detail, within the GHF basis, single-particle energies (SPEs) for GSM are obtained using the so-called $\hat{S}\text{-box}$ folded diagrams~\cite{CORAGGIO200543}, while the GSM effective interaction including the NO2B 3NF is derived using the $\hat{Q}\text{-box}$ folded diagrams~\cite{SUN2017227}. The $\hat{S}\text{-box}$ is by definition the one-body part of the $\hat{Q}\text{-box}$~\cite{SHURPIN197761,HU2020135206}. The $\hat{S}\text{-box}$ and $\hat{Q}\text{-box}$ folded-diagram method has been extended to the complex-$k$ space, which includes effects from the continuum and core polarization~\cite{SUN2017227,HU2020135206}. The complex-$k$ basis space is nondegenerate. Therefore, we use the nondegenerate EKK method~\cite{TAKAYANAGI201161} to process the $\hat{S}\text{-box}$ and $\hat{Q}\text{-box}$ calculations~\cite{SUN2017227,HU2020135206}. Due to the dramatic growth in the number of the GHF basis states with continuum partial waves included, the $\hat{Q}\text{-box}$ folded diagrams are calculated up to the second order, while the $\hat{S}\text{-box}$ folded diagrams are up to the third order.  In our previous work~\cite{HU2020135206}, it has been shown that the $\hat{S}\text{-box}$ up to the third order and $\hat{Q}\text{-box}$ up to the second order offer the good descriptions of nuclear states. The continuum effect enters into the model through both the effective interaction and active model space. With the SPEs and interaction renormalized, the effective shell-model Hamiltonian can be rewritten as
\begin{equation}
\begin{aligned}
\hat{H}_{\text{eff}}=&\sum_{p\in \mathrm{valence}}\epsilon_p\hat{a}_{p}^{\dagger} \hat{a}_{p}\\
&+\frac{1}{4} \sum_{pqrs\in \mathrm{valence}} (V^{\mathrm{NN}}_{pqrs}+W^{2 \mathrm{B}}_{pqrs})^{\text{eff}} \hat{a}_{p}^{\dagger} \hat{a}_{q}^{\dagger} \hat{a}_{s} \hat{a}_{r},
\end{aligned}
\end{equation}
where $\epsilon_p$ stands for the SPEs obtained by $\hat{S}\text{-box}$, and ``eff'' indicates the shell-model effective interaction renormalized to the valence space by the $\hat{Q}\text{-box}$  including the NO2B 3NF. 

The complex-symmetric GSM Hamiltonian~(6) is diagonalized in the model space using the Jacobi-Davidson method in the m-scheme~\cite{MICHEL2020106978}. Due to the presence of the nonresonant continuum, the matrix dimension grows dramatically when adding valence particles, which is a challenge for diagonalizing in the complex space. Therefore, we allow at most two valence particles in the continuum, which can give converged results~\cite{SUN2017227,HU2020135206,PRC70064313}.

\section{Results}
\begin{figure}[htb]
\includegraphics[width=0.48\textwidth]{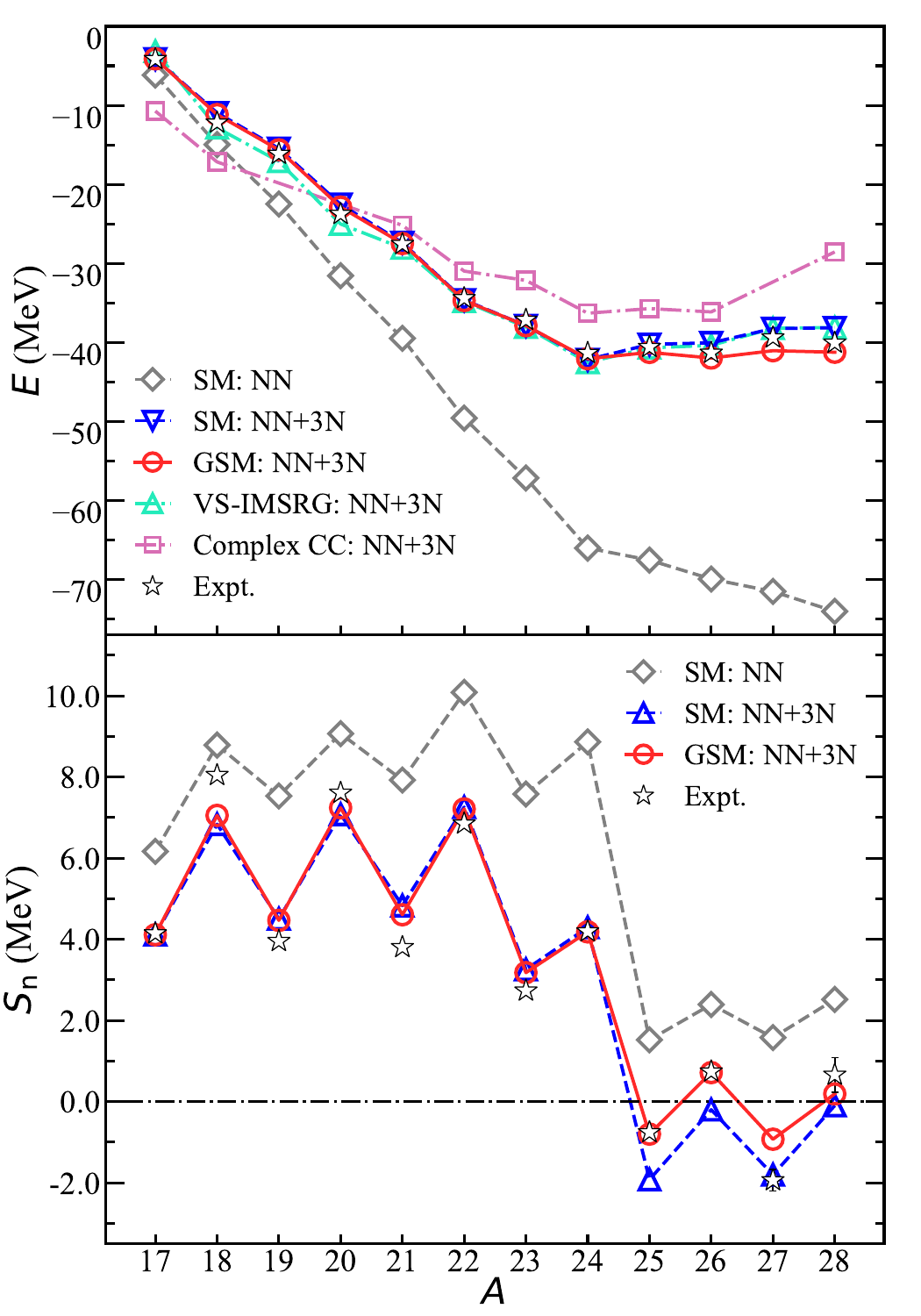}
\caption{\label{fig:eps1} $^{17\text{-}28}\text{O}$ g.s. energies (upper panel) with respect to the $^{16}\text{O}$ core, and the one-neutron separation energies $S_{\text n}$ (lower panel). ``NN" and ``NN+3N'' indicate calculations with NN-only and NN plus 3NF interactions, respectively. The experimental data of $^{17\text{-}26}$O have been available collected in AME2020~\cite{Wang_2021}, while $^{27,28}$O take the evaluations~\cite{Wang_2021}.} 
\end{figure}
Figure.~\ref{fig:eps1} shows the calculations of g.s. energies (upper panel) and one-neutron separation energies (lower panel) for neutron-rich $^{17\text{-}28}\text{O}$, compared with experimental data or evaluations. In order to see the effects of the 3NF and continuum coupling, we have also performed conventional real-energy shell-model (SM) calculations within the real-energy HF basis, with the same chiral interactions as those used in GSM. It is seen that the real-energy SM calculation based on the NN interaction gives overbound energies compared with data, and calculated one-neutron separation energies deviate significantly from data as well. It cannot give the correct position of the oxygen neutron dripline. The inclusion of 3NF improves remarkably the real-energy SM calculation, reproducing the dripline position at $^{24}$O. The inclusion of the continuum coupling (i.e., the GSM calculation) improves further the results in both binding energies and separation energies. The continuum effect becomes significant beyond the dripline $^{24}$O. The real-energy SM calculation with 3NF but without the continuum gives that $^{26}$O is more unbound than $^{25}$O, which disagrees with experimental data. The GSM calculations with both 3NF and continuum included give the correct unbound properties of $^{25}$O and $^{26}$O, showing that $^{25}$O is more unbound than $^{26}$O.

As shown in the lower panel of Fig.~\ref{fig:eps1}, $^{25,26}$O neutron separation energies are well described in the GSM calculations, better than in the real-energy SM calculations. In both calculations, 3NF is included. Experimental binding energies and neutron separation energies for $^{27,28}$O have not been available, but take the evaluations (AME2020)~\cite{Wang_2021}. The GSM calculation with 3NF included gives that $^{26}$O is stable against one-neutron emission, which is consistent with experiment, while the real-energy SM calculation without the continuum included shows that $^{26}$O is unstable against one-neutron emission. The experiment observed a two-neutron emission in $^{26}$O with a very small separation energy $S_{\text{2n}}$ of only -18 keV~\cite{PhysRevLett.116.102503}. The present GSM calculation provides $S_{\text{2n}}=-86$ keV, while the real-energy SM calculation with 3NF included gives $S_{\text{2n}}=-2.12$ MeV which is significantly different from the datum of $-18$ keV. The GSM calculation predicts that $^{28}$O should have no one-neutron emission, but two-neutron and four-neutron emissions are allowed.

In Fig.~\ref{fig:eps1}, we have also included the valence-space IMSRG (VS-IMSRG) calculations~\cite{PhysRevLett.118.032502} without the continuum effect considered. The results are very similar to the real-energy SM calculations (indicated by SM:NN+3N in Fig.~\ref{fig:eps1}). The complex coupled-cluster (complex CC) calculation takes the continuum effect into account, but with a density-dependent effective 3NF used~\cite{PhysRevLett.108.242501}. From these calculations compared, we should conclude that both 3NF and continuum coupling are important in the descriptions of weakly-bound and unbound nuclei. 

The $\nu 0d_{3/2}$ orbital is a resonance, which plays a special role in the resonant states of isotopes near $^{24}\text{O}$, particularly in odd-neutron oxygen isotopes around the neutron dripline. For example, the GSM calculation gives that the $3/2_{1}^{+}$ excited state in $^{23}\text{O}$ is dominated by a single-particle excitation from $\nu 1s_{1/2}$ to $\nu d_{3/2}$, with a calculated spectroscopic factor (SF) $\left[^{22} \mathrm{O}_{\mathrm{g}.\mathrm{s}.}\otimes \nu d_{3/2}\right]^{3/2^{+}}$ of 0.95. The unbound $^{25}\text{O}$ resonant ground state has the odd neutron occupying the $\nu d_{3/2}$ orbital, with a calculated SF $\left[^{24}\mathrm{O}_{\mathrm{g}.\mathrm{s}.}\otimes \nu d_{3/2}\right]^{3/2^{+}}$ of 0.95. With the $\nu d_{3/2}$ resonance and continuum considered, the GSM calculation provides that the excitation energy ($E_{\text {ex}}$) of the $^{23}\text{O}$ $3/2_{1}^{+}$ state is 4.45 MeV, which agrees well with the datum of 4.00(2) MeV~\cite{PhysRevLett.98.102502}, while the real-energy SM calculation gives $E_{\text {ex}}=5.67$ MeV. The GSM can also well describe the experimental one-neutron separation energy of the $^{25}\text{O}$ ground state, see Fig.~\ref{fig:eps1}.

\begin{figure}[htb]
\includegraphics[width=0.48\textwidth]{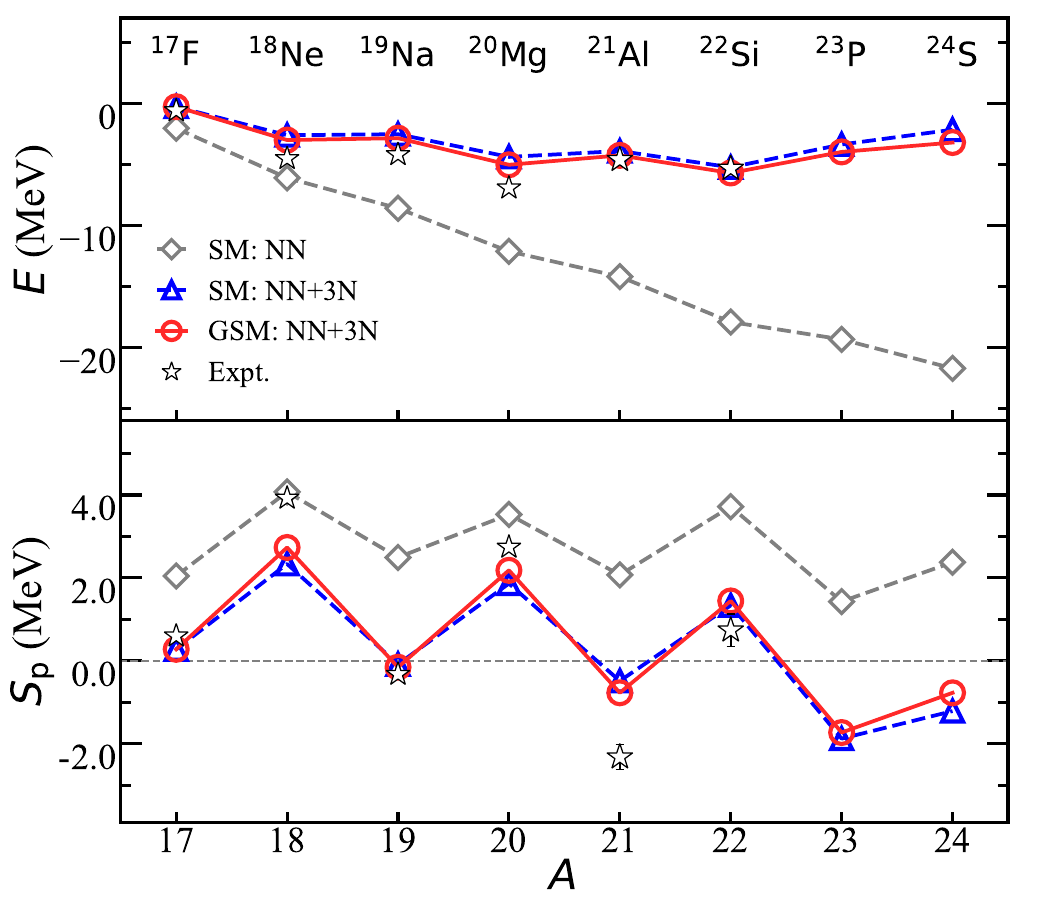}
\caption{\label{fig:eps2} Similar to Fig.~\ref{fig:eps1}, but for $N=8$ isotones and the one-proton separation energy ($S_\text{p}$). Experimental data for $A\leq 20$ isotones have been available~\cite{Wang_2021}, while evaluations~\cite{Wang_2021} are taken for $^{21}$Al and $^{22}$Si.}
\end{figure}

As mirror partners of the oxygen isotopes with respect to $N = Z$, proton-rich $N=8$ isotones are somewhat different on account of the presence of Coulomb interaction. Here, the Coulomb repulsion results in a low particle-emission threshold, while the additional Coulomb barrier hinders the diffusion of proton wave functions. We show in Fig.~\ref{fig:eps2} the calculated g.s. energies and one-proton separation energies for $N=8$ isotones in the GSM, compared with real-energy SM calculations based on the same chiral interactions. Similar to the $Z=8$ mirror partners in Fig.~\ref{fig:eps1}, real-energy SM calculations with NN-only lead to overbound g.s. energies. The overbinding is significantly improved by including 3NF, as shown in Fig.~\ref{fig:eps2}. We notice that the difference is small between the SM and GSM calculations when 3NF is included, see Fig.~\ref{fig:eps2}, which indicates a weak continuum effect in the isotones. This is because in proton-rich nuclei the Coulomb barrier weakens the coupling to the continuum. Starting from $^{19}\text{Na}$, the odd-proton $N=8$ isotones become unbound with proton emissions. This is well reproduced in both SM and GSM calculations with 3NF included. The unbound property of $^{19}\text{Na}$ makes $^{20}\text{Mg}$ a Borromean structure, which is required for the successive proton capture $^{18}\text{Ne}(2p,\gamma)^{20}\text{Mg}$ to bridge the waiting point in the astrophysical $rp$-process~\cite{PhysRevC.51.392}. 

$^{22}\text{Si}$ is an interesting nucleus for which the experimental mass has not been available, but the evaluation has been given in~\cite{Wang_2021}. Current experiments by indirect mass measurements gave very different estimations of the $^{22}\text{Si}$ two-proton separation energy, $S_{2p}\approx -108(125)$ keV in~\cite{XU2017312} and $S_{2p}\approx 645(100)$ keV in~\cite{babo:tel-01461303}. The SM~calculation ~\cite{PhysRevLett.110.022502} suggested that $^{22}\text{Si}$ is a candidate of two-proton decay, while the GSM calculation with a phenomenological potential~\cite{PhysRevC.100.064303} gave a weakly bound $^{22}\text{Si}$. The present GSM calculation predicts a two-proton separation energy of 674 keV for $^{22}\text{Si}$. High-resolution experimental direct measurements are required for the possible proton-dripline nucleus.
\begin{figure}[htb]
\includegraphics[width=0.48\textwidth]{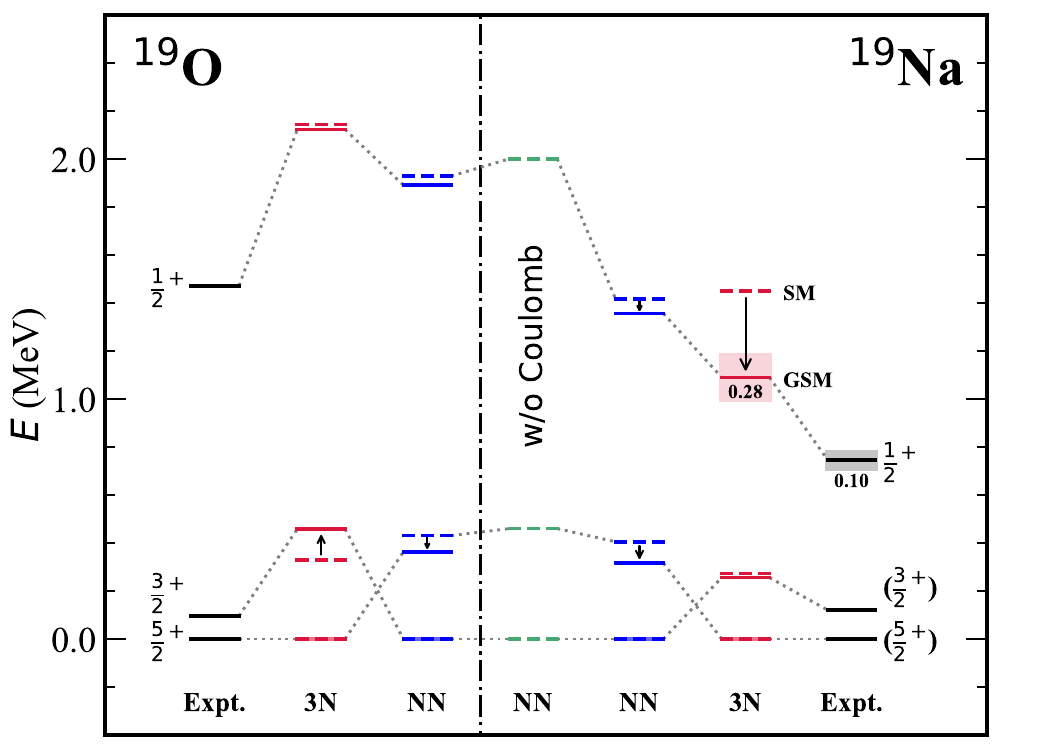}
\caption{\label{fig:eps3} Excitation spectra of $A=19$ mirror partners, $^{19}\text{O}$ and $^{19}\text{Na}$. ``NN'' and ``3N'' indicate calculations with NN-only and 3NF included, respectively. Dashed levels present SM calculations without continuum included, while solid levels give GSM calculations with continuum included. Shading indicates the resonance of the $1/2^+$ TES state with width (in MeV) given below the level. Data are from~\cite{ensdf} for $^{19}\text{O}$ and~\cite{ensdf,PhysRevC.82.054315,PhysRevC.67.014308} for $^{19}\text{Na}$.}
\end{figure} 

As mentioned above, the TES phenomenon that may happen in mirror nuclei is a reflection of different asymptotic behaviors of states~\cite{PhysRev.88.1109,PhysRev.81.412}. Therefore, the continuum coupling would play a role in TES. For this mass region which we are investigating, the experiment~\cite{PhysRevC.67.014308} has observed a pronounced TES in the $1/2^{+}$ excited state between the mirror partners $^{19}\text{Na}$ and $^{19}\text{O}$. The SM calculation with an empirical interaction~\cite{PhysRevC.89.044327} has analyzed this TES, showing a requirement of readjusting interaction matrix elements to reproduce the observed TES. Here, we investigate the 3NF and continuum effects on the TES, in the framework of the self-consistent {\it ab initio} GSM with the chiral interactions used above. Figure~\ref{fig:eps3} shows the excitation spectra of the mirror partners $^{19}\text{Na}$ and $^{19}\text{O}$. Calculations with NN-only results in a $3/2^{+}$ g.s.  instead of the experimental $5/2^{+}$ g.s. for both $^{19}\text{Na}$ and $^{19}\text{O}$. When 3NF is included, the correct ordering between the $5/2^{+}$ g.s. and $3/2^{+}$ excited state is obtained. 

For the proton-rich $^{19}\text{Na}$, we have also performed a calculation with the Coulomb interaction excluded, because the Coulomb energy difference is considered to be the main factor leading to the mirror symmetry breaking in excitation spectra~\cite{PhysRev.88.1109,PhysRev.81.412}. Indeed, we see that in Fig.~\ref{fig:eps3} a good mirror symmetry appears in the excitation spectra if the Coulomb interaction is excluded. This also indicates that effects from the charge symmetry breaking (CSB) and the charge independence breaking (CIB)~\cite{MACHLEIDT20111} are small. When the Coulomb potential is included, the $^{19}\text{Na}$ $1/2^+$ energy drops significantly, while the spacing between $5/2^+$ and $3/2^+$ levels remains nearly unchanged (see Fig.~\ref{fig:eps3}). The $^{19}\text{Na}$ $1/2^+$ state has almost a pure proton $(0d_{5/2})^{2}(1s_{1/2})^{1}$ configuration. The lack of a centrifugal barrier in the $1s_{1/2}$ orbital causes the leakage of the wave function, and leads to a smaller Coulomb repulsion energy compared with the $5/2^+$ and $3/2^+$ states, thus makes a drop of the $1/2^+$ level.

As mentioned above, 3NF is important to reproduce the correct ordering between the $3/2^+$ and $5/2^+$ states (see the second column from the right in $^{19}$Na). The 3NF does not seem to significantly affect the ${1/2}^+$ level in the real-energy SM calculation (dashed), but in fact it increases remarkably the spacing between the ${1/2}^+$ and $5/2^+$ states, which means an increase in the ${1/2}^+$ excitation energy with taking the $5/2^+$ state as the ground state. The $^{19}$Na g.s. and excited states are bound in calculations with NN-only, while they become unbound when 3NF is included, see also Fig.~\ref{fig:eps2}. The GSM calculation with 3NF included gives very narrow resonances of $^{19}$Na $5/2^+$ g.s. and $3/2^+$ excited state, which agrees with the experimental estimation of a narrow resonant g.s.~\cite{PhysRevC.82.054315}. The experiment ~\cite{PhysRevC.67.014308} showed a resonant $1/2^+$ excited state with a width of 0.10 MeV. Therefore, the continuum effect should be significant in the $1/2^+$ state. Indeed, we see that the continuum coupling lowers significantly the $1/2^+$ level in the GSM calculation, which makes the calculated excitation energy closer to the datum~\cite{PhysRevC.67.014308}. The calculated width of 0.28 MeV also agrees reasonably with the experimental width of 0.10 MeV~\cite{PhysRevC.67.014308}. We see that the complex-energy GSM calculation well describes the observed TES in the $1/2^+$ state, in which both 3NF and continuum play important roles.

\begin{figure}[htb]
\includegraphics[width=0.48\textwidth]{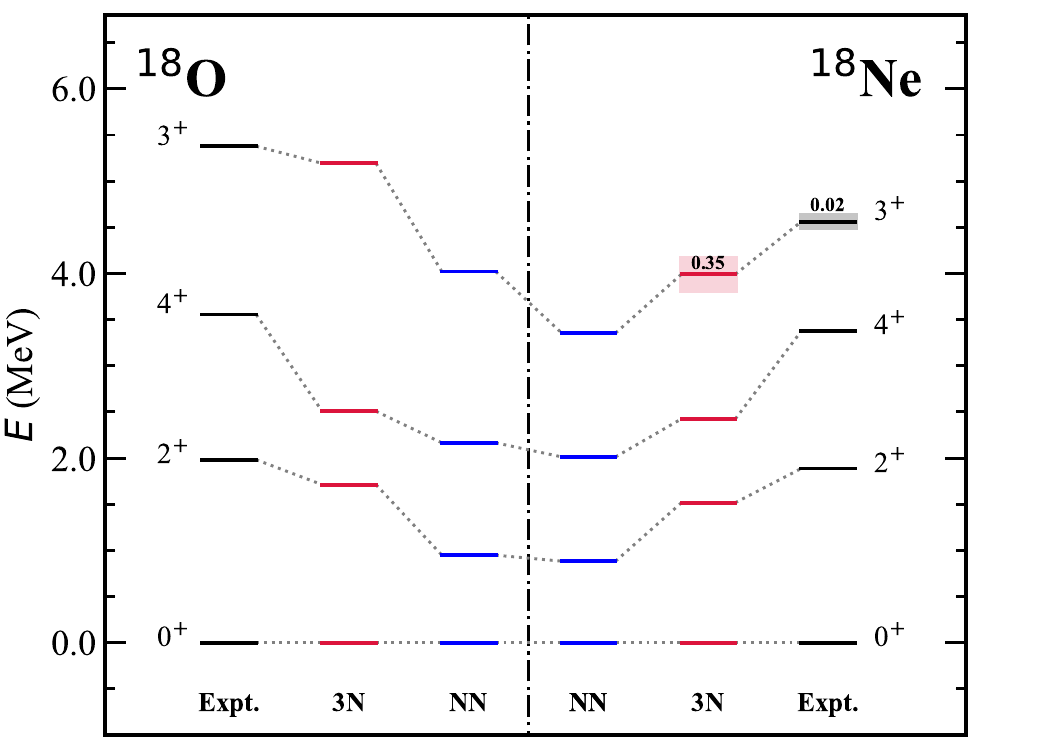}
\caption{\label{fig:eps4} GSM calculations with NN only and with 3NF included (3N) for the mirror partners of $^{18}\text{O}$ and $^{18}\text{Ne}$, compared with data~\cite{ensdf,PhysRevLett.83.45}}
\end{figure}

Figure~\ref{fig:eps4} shows the excitation spectra for another pair of mirror nuclei of $^{18}\text{O}$ and $^{18}\text{Ne}$ in which the TES was observed experimentally in the excited $3^+$ state~\cite{PhysRevLett.83.45}. The resonant $3^+$ state plays an important role in the key reaction $^{17}$F($p,\gamma$)$^{18}$Ne of stellar explosions~\cite{PhysRevLett.83.45}. The presence and its property of the $3^+$ state can change the reaction rate. Indeed, the calculation well shows the TES in the $3^+$ level. The GSM calculation gives a resonant $3^+$ state with an almost pure proton $0d_{5/2}\otimes 1s_{1/2}$ configuration in which the $1s_{1/2}$ orbital is a strong $l=0$ resonance. The compressed spectra are significantly improved when 3NF is included in calculations. 

\begin{figure}[htb]
\includegraphics[width=0.48\textwidth]{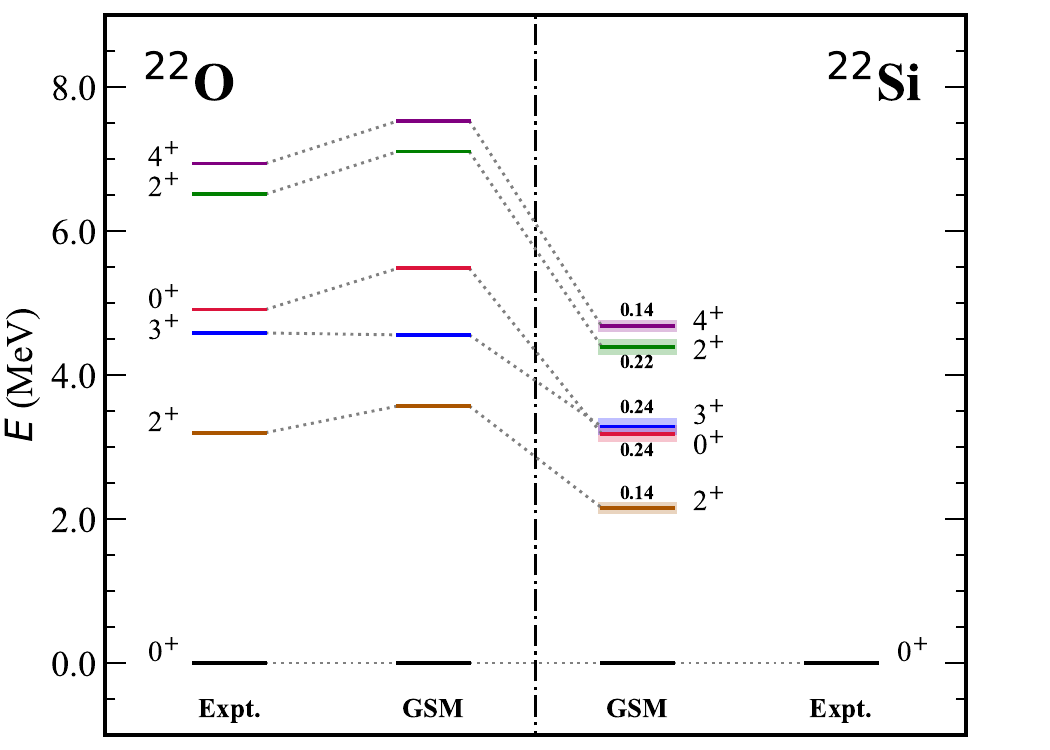}
\caption{\label{fig:eps5} GSM calculations of spectra with 3NF included, for the mirror nuclei $^{22}\text{O}$ and $^{22}\text{Si}$. The experimental data of $^{22}\text{O}$ is from~\cite{ensdf}.}
\end{figure}

In Fig.~\ref{fig:eps5}, we predict the excitation spectra of the mirror partners $^{22}\text{O}$ and $^{22}\text{Si}$. $^{22}$Si is predicted to be the proton dripline of the $N=8$ isotonic chain. There has been no excited state observed in $^{22}$Si. Our calculation gives that $^{22}$Si has a bound g.s. but unbound resonant excited states. In the g.s. the six valence protons outside the $^{16}$O core occupy mainly the $0d_{5/2}$ orbital which is bound, while in excited states one or two valence protons are excited to the $1s_{1/2}$ orbital which is a resonance. The $s_{1/2}$ orbital has no centrifugal barrier due to $l=0$, thus the wave function of the state containing a significant $s_{1/2}$ component can more spread in space, which leads to a TES in the level. Indeed, we see that all the calculated $^{22}$Si levels given in Fig.~\ref{fig:eps5} show TES. This predicts an interesting candidate for future experiments, with multiple low-lying TES levels in a nucleus. 

\section{Conclusions}
With both three-nucleon force and continuum coupling included, we have developed the complex-energy {\it ab initio} Gamow shell model with a core. The Berggren representation is used for the Gamow shell model, which presents bound, resonance and nonresonant continuum states on equal footing. The Berggren basis is generated by the complex-energy Gamow Hartree-Fock method using the same interaction as that used in the Gamow shell model. Starting from the chiral two-nucleon (N$^3$LO) and three-nucleon (N$^2$LO) forces, the realistic effective Hamiltonian of the Gamow shell model is established using the many-body perturbation theory (named $\hat{S}\text{-}$ and $\hat{Q}\text{-box}$ diagrams) in the complex space. For neutron-rich $Z=8$ isotopes and proton-rich $N=8$ isotones, the shell model chooses $^{16}$O as the core, and takes the model space of \{$1s_{1/2}, 0d_{5/2}, 0d_{3/2}, 1p_{3/2}, 0f_{7/2}$\} and plus their continuum partial waves of the resonances in the Gamow shell model. For shell-model calculations, the three-nucleon interaction is approximated to two-body level by the normal-ordering method.

$Z(N)=8$ isotopes (isotones) have been investigated. We find that the calculations with two-nucleon interaction only cannot reasonably describe binding energies, nucleon separation energies and excitation spectra. The inclusion of three-nucleon force can significantly improve the calculations. For example, the calculation with three-nucleon force included can give the correct ground states of the $A=19$ mirror nuclei $^{19}$O and $^{19}$Na. The inclusion of the continuum coupling improves further the results. The $^{26}$O (e.g., its one- and two-neutron separation energies) is better described in the calculation with the continuum effect considered. With both three-nucleon force and continuum coupling included, the dripline positions can be reproduced. 

The Thomas-Ehrman shift observed in the excitation levels of mirror nuclei was suggested mainly due to different Coulomb energies between states. In the states containing a significant $s_{1/2}$ component, the wave functions can more spread in space due to no centrifugal barrier of the $l=0$ orbital, and the continuum coupling is strong. Indeed, the Thomas-Ehrman shift was observed mainly in excited states around or above the threshold of particle emission, in which the continuum effect can be significant. In $^{19}$Na, the Gamow shell-model calculation with both three-nucleon force and continuum coupling included gives a resonant $1/2^+$ excited state, which agrees with data, and the observed Thomas-Ehrman shift is well described. For this $1/2^+$ state, three-nucleon force and continuum coupling produce a combined effect on its Thomas-Ehrman shift. In more detail, the asymptotic behavior of the wave function of a weakly-bound or unbound state is accurately treated in the Gamow shell model, but the final result also depends on the energy which is affected by the three-nucleon interaction.

As a prediction, we have investigated excitation spectra for the heavier mirror partners of $^{22}$O and $^{22}$Si. In $^{22}$Si, all the calculated excited states are unbound with proton emissions. They contain the significant $s_{1/2}$ component, and hence the Thomas-Ehrman shift is seen in the low-lying excitation levels of $^{22}$Si. This predicts a special case with multiple Thomas-Ehrman shift levels in one pair of mirror nuclei, which would be interesting for future experiments. 

\section{Acknowledgements}
Discussions with N. Michel, S.M. Wang and Z.H. Sun are gratefully acknowledged. The complex-symmetric Hamiltonian is diagonalized using the GSM code provided by N. Michel, which is publicly available on \href{https://github.com/GSMUTNSR}{https://github.com/GSMUTNSR}. This work has been supported by the National Key R\&D Program of China under Grant No. 2018YFA0404401; the National Natural Science Foundation of China under Grants No. 11835001, No. 11921006 and No. 12035001; China Postdoctoral Science Foundation under Grant No. BX20200136; the State Key Laboratory of Nuclear Physics and Technology, Peking University under Grant No. NPT2020KFY13; and the CUSTIPEN (China-U.S. Theory Institute for Physics with Exotic Nuclei) funded by The U.S. Department of Energy, Office of Science under Grant No. de-sc0009971. We acknowledge the High-Performance Computing Platform of Peking University for providing computational resources.



\bibliographystyle{elsarticle-num_noURL}
\bibliography{article}





\end{document}